\documentclass{article}%
\usepackage{amsfonts}
\usepackage{makeidx}
\usepackage{amsmath}
\usepackage{amssymb}
\usepackage{graphicx}%
\providecommand{\U}[1]{\protect\rule{.1in}{.1in}}
\newtheorem{theorem}{Theorem}

\newtheorem{proposition}[theorem]{Proposition}

\begin{document}

\title{Quadratic conservation laws and collineations: a discussion}
\author{Leonidas Karpathopoulos\\{\ \ \textit{Faculty of Physics, Department of
Astronomy-Astrophysics-Mechanics,}}\\{\ \textit{University of Athens, Panepistemiopolis, Athens 157 83, Greece}}
\and Michael Tsamparlis\thanks{Email: mtsampa@phys.uoa.gr}\\{\textit{Faculty of Physics, Department of Astronomy-Astrophysics-Mechanics,}}\\{\ \textit{University of Athens, Panepistemiopolis, Athens 157 83, Greece}\ }
\and Andronikos Paliathanasis\thanks{Email: anpaliat@phys.uoa.gr}\\{\ \textit{Instituto de Ciencias F\'{\i}sicas y Matem\'{a}ticas, }}\\{\textit{Universidad Austral de Chile, Valdivia, Chile}}\\{\textit{Institute of Systems Science, Durban University of Technology}}\\{\textit{Durban 4000, Republic of South Africa}}}
\date{}
\maketitle

\begin{abstract}
Every second order system of autonomous differential equations can be
described by an autonomous holonomic dynamical system with a Lagrangian part,
an effective potential and a set of generalized forces. The kinematic part of
the Lagrangian defines the kinetic metric which subsequently defines a
Riemannian geometry in the configuration space. We consider the generic
function $I=K_{ab}(t,q^{c})\dot{q}^{a}\dot{q}^{b}+K_{a}(t,q^{c})\dot{q}%
^{a}+K(t,q^{c})$ and require the quadratic first integral condition $dI/dt=0$
without involving any type of symmetry Lie or Noether. Condition $dI/dt=0$
leads to a system of equations involving the coefficients $K_{ab}%
(t,q^{c}),K_{a}(t,q^{c}),K(t,q^{c})$ whose solution will produce all possible
quadratic first integrals of the original system of autonomous differential
equations. We show that the new system of equations relates the quadratic
first integrals of the holonomic system with the geometric collineations of
the kinetic metric and in particular with the Killing tensors of order two. We
consider briefly various results concerning the Killing tensors of
second-order and prove a general formula which gives in the case of a flat
kinetic metric the generic Killing tensor in terms of the vectors of the
special projective algebra of the kinetic metric. This establishes the
connection between the geometry defined by the kinetic metric and the
quadratic first integrals of the original system of differential equations.

\end{abstract}


PACS - numbers: 2.40.Hw, 4.20.-q, 4.20.Jb, 04.20.Me, 03.20.+i, 02.40.Ky

\section{Introduction}

\label{Introduction}

At the end of the 19th century Sophus Lie published his work on the theory of
transformation groups \cite{lie1,lie2,lie3}. In particular he introduced the
idea of transformations which leave invariant functions and geometric objects.
The application of Lie's method has led to new interesting results is the area
of differential equations.

A Lie symmetry of a differential equation generated by the vector
$X=\xi(t,q,\dot{q})\partial_{t}+\eta^{a}(t,q,\dot{q})\partial_{q^{a}}\ $ is\ a
point transformation in the jet space $J^{1}\left\{  t,q^{a},\dot{q}\right\}
$ which preserves the form of the equation and transforms solutions into
solutions. In this work we are interested on dynamical systems of second-order
differential equations of the form
\begin{equation}
\ddot{q}^{a}=\omega^{a}(t,q,\dot{q}). \label{FL.0}%
\end{equation}
Eqn. (\ref{FL.0}) defines the Hamiltonian vector field $\Gamma$ in the jet
space $J^{1}\left\{  t,q^{a},\dot{q}\right\}  $ as follows%
\begin{equation}
\Gamma=\frac{d}{dt}=\frac{\partial}{\partial t}+\dot{q}^{a}\frac{\partial
}{\partial q^{a}}+\omega^{a}\frac{\partial}{\partial\dot{q}^{a}}. \label{FI.3}%
\end{equation}

The mathematical condition for a vector field $X^{[1]}=\xi(t,q,\dot
{q})\partial_{t}+\eta^{a}(t,q,\dot{q})\partial_{q^{a}}+\left(  \dot{\eta}%
^{a}-\dot{q}^{a}\dot{\xi}\right)  \partial_{\dot{q}^{a}}\ $ in $J^{1}\left\{
t,q^{a},\dot{q}\right\}  $ to be a Lie symmetry of (\ref{FL.0}) is that there
exists a function $\lambda\left(  t,q,\dot{q}\right)  $ such that
\begin{equation}
\lbrack X^{\left[  1\right]  },\Gamma]=\lambda(t,q,\dot{q})\Gamma.
\label{FL.0.1}%
\end{equation}

It can be shown that the set of Lie symmetries of (\ref{FL.0}) span a Lie
algebra. If the point transformation is in the base space $\{t,q^{a}\}$ that
is $\xi(t,q),\eta^{a}(t,q)$ the Lie symmetry is called a point Lie symmetry.
If the point transformation is in the jet space $J^{1}\left\{  t,q^{a},\dot
{q}\right\}  $ it is called a dynamical Lie symmetry. Usually with the term
Lie symmetry one refers to the point Lie symmetries$.$ However in the
literature there have been considered many types of dynamical Lie symmetries
e.g. contact symmetries, nonclassical symmetries, invariant surface symmetries
and many others, (see \cite{sym1,sym2,sym3,sym4,sym5} and references therein).
Furthermore in dynamical Lie symmetries one has an extra degree of freedom
which is removed if one demands an extra condition in which case one works
with the so called gauged dynamical symmetries. In this respect one usually
requires the gauge condition $\xi=0$ so that the generator is simplified to
$X^{[1]}=\eta^{a}(t,q,\dot{q})\partial_{q^{a}}+\Gamma(\eta^{a})(t,q,\dot
{q})\partial_{\dot{q}^{a}}.$ This gauge condition we shall assume in the following.

In case the dynamical equations (\ref{FL.0}) follow from a Lagrangian
$L\left(  t,q,\dot{q}\right)  ,$ then Emmy Noether \cite{noe1}
showed\footnote{For an alterantive proof of the Emmy Noether's first theorem
we refer the reader to \cite{klev}.} that if a Lie symmetry of (\ref{FL.0})
leaves in addition the action integral $S=\int L\left(  t,q,\dot{q}\right)
dt$ invariant, that is $X\left(  S\right)  =\lambda S,$ then there exists a
function $\Phi\left(  t,q,\dot{q}\right)  $ given by the expression%
\begin{equation}
\Phi\left(  t,q,\dot{q}\right)  =\xi\left(  \dot{q}^{a}\frac{\partial
L}{\partial\dot{q}^{a}}-L\right)  -\eta^{a}\frac{\partial L}{\partial\dot
{q}^{a}}+f \label{FI.4}%
\end{equation}
such that $\Gamma\left(  \Phi\right)  =0,$ that is, $\Phi$ is a first integral
of (\ref{FL.0}). Such Lie symmetries are called Noether symmetries. For a
point Noether symmetry we have the additional property $X(\Phi)=\lambda\Phi$
which means that $\Phi$ is an invariant first integral. The condition for the
existence of a point Noether symmetry is \cite{Sarlet,SarletCantrijn 81}
\begin{equation}
X^{[1]}L+\frac{d\xi}{dt}L=\frac{df}{dt}. \label{FI.1}%
\end{equation}
Condition (\ref{FI.1}) is called the Noether Condition$.$ Noether point
symmetries span a subalgebra of the (finite dimensional) Lie algebra of point
Lie symmetries.

The function $f$ in (\ref{FI.1}) usually is refereed as \textquotedblleft
gauge\textquotedblright\ function. This is an incorrect terminology. The
function $f$ is a boundary term introduced to allow for the infinitesimal
changes in the value of the Action Integral produced by the infinitesimal
change in the boundary of the domain caused by the transformation of the
variables in the Action Integral\footnote{An interesting discussion on\ the
Noether's theorem which clarifies various misinterpretations in the
litterature can be found in \cite{leachnoe1}.
\par
{}}. Because the dynamical equations which interest us mostly are of second
order we are mainly interested for quadratic first integrals.

The main use of Noether symmetries is to provide the Noether integrals which
are used to simplify the system of equations and, if there are enough of them,
even to determine the solution. Necessary conditions for a Noether point
symmetry are a. that there exists a Lagrangian function $L\left(  t,q,\dot
{q}\right)  $ which describes the field equations, and \ b. that conditions
(\ref{FI.1}) are satisfied for the given Lagrangian. However, the definition
of a Lagrangian for a given dynamical system is not unique.\ That is, it is
possible that there exist non-equivalent (i.e. not differing by a perfect
differential) Lagrangians describing the same dynamical equations which have
different Noether symmetries (see \cite{jlm1,jlm2,jlm3}). Therefore it is
clear that when we refer to a Noether symmetry of a given dynamical system we
should mention always the Lagrangian function we are assuming. More recently
there have been proposed other approaches
\cite{hojman,Crampin,prince,Leach08,ll1,vuja,vuja2,moyo} to determine first
integrals of dynamical equations. Some of the authors call the resulting first
integrals non-Noetherian.

For instance, the quadratic first integral of the Kepler problem, the
Runge-Lenz vector, can be constructed by the first-order invariant of a scale
symmetry vector \cite{prince}, while Crampin in \cite{Crampin} calls the
Runge-Lenz vector a non-Noetherian hidden symmetry. However the
characterization of the Runge-Lenz vector as non-Noetherian conservation law
is not accurate. The Runge-Lenz vector cannot be constructed directly from
formula (\ref{FI.4}) with the use of point Noether symmetries for the
classical Lagrangian of the Kepler problem. However it is derived if one uses
instead contact symmetries \cite{kalotas}. Another classical example is the
Lewis invariant of the Ermakov-Pinney system \cite{lew1,lew2,lew3} where there
are more than one approaches to reach the same result. The Runge-Lenz and the
Lewis invariant differ from the invariants of the (angular) momentum in the
sense that they are quadratic conservation laws.

Our approach is summarized as follows. Consider a function $I(t,q^{a},\dot
{q}^{a})$ which is linear and quadratic in the velocities with coefficients
which depend only on the coordinates $t,q^{a}.$ Demand that $I$ satisfies the
condition of being a first integral, that is,%
\begin{equation}
\frac{dI}{dt}=0. \label{DS1.10a}%
\end{equation}
Using the dynamical equations determine the resulting conditions for the
coefficients of $I$ whose solution will give all possible quadratic first
integrals of this form of the given dynamical equations. In case the dynamical
equations follow from a regular Lagrangian define the kinetic metric
$\gamma_{ij}=\frac{\partial^{2}L}{\partial\dot{q}^{i}\dot{q}^{j}}$ and
consider the gauge $\xi=0$ and compute the remaining component $\eta^{i}$ of
the generator of the point transformation from the Cartan condition%

\begin{equation}
\eta^{i}(t,q^{i},\dot{q}^{i})=-\gamma^{ij}\frac{\partial I_{2}}{\partial
\dot{q}^{j}}. \label{DS1.12}%
\end{equation}
Then the symmetry vector which generates the Noether symmetry which admits
this first integral is%
\begin{equation}
X^{[1]}=\eta^{i}\frac{\partial}{\partial q^{i}}+\mathbf{\Gamma}(\eta^{i}%
)\frac{\partial}{\partial\dot{q}^{i}} \label{DS1.15}%
\end{equation}
where $\Gamma$ is the Euler vector field introduced in (\ref{FI.3}). Because
we can always consider the Lagrangian to be the kinetic energy, this
interpretation of a first integral in terms of an associated Noether symmetry
appears to be always possible. As will be shown our result generalizes those
of \cite{p1,p2} for point symmetries and \cite{Crampin,kalotas} for a family
of quadratic first integrals.

A\ further point we shall consider is the relation between the symmetries
generating the quadratic integrals and the symmetries of the geometry of the
kinetic metric defined by the dynamical system. This metric defines a geometry
in the solution space which admits certain geometric symmetries which are
called collineations. It is natural one to expect that there must be a
relation between the Lie and the Noether symmetries of the dynamical equations
and the collineations of the kinetic metric.

Early studies of this problem have been done by Katzin \& Levine
\cite{katz1,katz2}, Aminova \cite{ami1,ami2} and others. A\ more recent study
which gives the complete answer for the case of point symmetries has been done
in \cite{p1,p2} where it has been shown that the Lie point symmetries of the
dynamical equations are given by the special projective algebra of the kinetic
metric and the Noether point symmetries by the Homothetic algebra of the
kinetic metric. Such a firm geometric result does not exist for the dynamical
Lie/Noether symmetries. \ Similar results have been derived for second-order
partial differential equations \cite{p3,p4,p5,p5a} and for regular
perturbative Lagrangians \cite{p6}. Furthermore results for singular
Lagrangians which support constrained dynamical systems can be found in
\cite{p7,p8}.

The plan of the paper is as follows. In Section \ref{sec2} are given the basic
properties of collineations and Killing tensors of Riemannian manifolds. Some
new results are given which enable one to construct Killing tensors of order 2
from the collineations. In section \ref{sec3} we discuss the relation between
collineations of the kinetic metric and quadratic first integrals. Finally, in
Section \ref{sec5} we discuss our results and draw our conclusions.

\section{Collineations and Killing tensors of second-order}

\label{sec2}

Consider a Riemannian space $V^{n}$ with metric tensor $g_{ab}.$ Let
$\mathbf{\Omega}\left(  x^{c}\right)  $ be a linear geometric
object\footnote{With the term \textquotedblleft geometric
object\textquotedblright\ we mean a quantity in $V^{n}$ associated with a
transformation rule. Not necessarily a tensor field.} in $V^{n}$ which under
the action of the infinitesimal transformation $x^{a^{\prime}}=x^{a^{\prime}%
}\left(  x^{b},\varepsilon\right)  $ transforms as follows\
\begin{equation}
\mathbf{\bar{\Omega}}\left(  \mathbf{\Omega},x^{c},x^{c^{\prime}}\right)
=J\left(  x^{c},x^{c^{\prime}}\right)  ...J\left(  x^{c},x^{c^{\prime}%
}\right)  \mathbf{\Omega}+\mathbf{\Xi}\left(  x^{c},x^{c^{\prime}}\right)  ,
\end{equation}
where $J\left(  x^{c},x^{c^{\prime}}\right)  $ denotes the Jacobian matrix of
the transformation.

A collineation of the geometric object $\mathbf{\Omega}\left(  x^{c}\right)  $
under the action of the infinitesimal transformation $x^{a^{\prime}%
}=x^{a^{\prime}}\left(  x^{b},\varepsilon\right)  $ generated by the vector
field $X\left(  x^{c}\right)  $ is defined by%
\begin{equation}
\mathbf{\Psi}\left(  x^{c}\right)  =\lim_{\varepsilon\rightarrow0}%
\frac{\mathbf{\bar{\Omega}}\left(  \mathbf{\Omega,}x^{a^{\prime}}\left(
x^{b},\varepsilon\right)  \right)  \mathbf{-\Omega}\left(  x^{b}\right)
}{\varepsilon}. \label{ss.01}%
\end{equation}
where $\Psi$ has the same numbers of components and symmetries of the indices
with $\mathbf{\Omega}.$The rhs of the latter expression is the definition of
the Lie derivative~wrt $X$ \ therefore (\ref{ss.01}) is written in the simple
form $\mathcal{L}_{X}\mathbf{\Omega}=\mathbf{\Psi.}$

Collineations involving geometric objects derived from the metric tensor,
$g_{ab}$, i.e. the Christofell symbols $\Gamma_{bc}^{a}$, the Ricci tensor
etc. play an important role in the mathematical structure of the Riemannian
manifold $V^{n}$ and on the properties of the physical systems defined on
$V^{n}.$

\subsection{Collineations concerning the metric tensor}

In the present work of special interest are two families of geometric
collineations. The first family are the conformal motions defined by the
requirement
\begin{equation}
\mathcal{L}_{X}g_{ab}=2\psi\left(  x^{c}\right)  g_{ab}%
\end{equation}
in which the vector field $X$ is called a Conformal Killing Vector (CKV) and
the function $\psi(x^{c})$ the conformal factor. One shows easily that
$\psi=\frac{1}{\dim V^{n}}X_{~~;c}^{c}$. In case $\psi_{;ij}=0,$ $\psi_{,i}=0$
with $\psi\neq0$ and $\psi=0$ the CKV specializes to a special CKV, a
homothetic vector (HV) and to a Killing vector (KV) respectively.

The second family are the projective collineations (PC) defined by the
requirement%
\begin{equation}
L_{X}\Gamma_{bc}^{a}=2\delta_{(b}^{a}\phi,_{c)}, \label{aa.01}%
\end{equation}
where $\phi\left(  x^{c}\right)  $ is a general function. In case $\phi
_{;ij}=0$ with $\phi_{,i}\neq0$ and $\phi=0$ the PC\ specializes to a special
PC and an Affine collineation (AC) respectively. An AC which is not a KV or
the HV is called a proper AC.

Using the identity $L_{\eta}\Gamma_{bc}^{a}=\eta_{;bc}^{a}-R_{bcd}^{a}\eta
^{d},~$one rewrites condition (\ref{aa.01}) as follows
\begin{equation}
X_{;(bc)}^{a}-2\delta_{(b}^{a}\phi,_{c)}=R_{bcd}^{a}X^{d}. \label{FL9.a.2}%
\end{equation}
from which follows that
\begin{equation}
X_{(a;b}-2g_{ab}\phi)_{;c)}=0.~ \label{aa.2}%
\end{equation}

A Killing tensor (KT) $K$ is defined by the requirement%
\begin{equation}
\left[  K,g_{ab}\right]  _{SN}=0
\end{equation}
or equivalently
\begin{equation}
K_{\left(  a_{1}...a_{k};c\right)  }=0
\end{equation}
where $\left[  ,\right]  _{SN}$ denotes the Schouten-Nijenhuis bracket, a
generalization of the Lie derivative. If $K$ is a vector field then $\left[
,\right]  _{SN}$ reduces to the Lie derivative and the KT reduces to a KV.
From (\ref{aa.2}) it follows that the second-rank tensor $C_{ab}=X_{\left(
a;b\right)  }-2g_{ab}\phi$ is a Killing tensor of rank two. Therefore it is
clear that PCs can be used to construct KTs.

In the special case that the KT $C_{ab}$ comes form a vector field $L^{a}$
such that $C_{ab}=L_{(a;b)},$ we can define the vector field $M_{a}%
=L_{a}-X_{a},~$where $X^{a}$ is a PC which satisfies the relation%
\[
M_{(a;b)}=-2\phi g_{ab},
\]
that is, $M^{a}$ is a CKV with conformal factor $-\phi.$

Therefore if in a space there are $m$ CKVs $M^{a}$ and $m$ PCs $X^{a}$ such
that the conformal factor is minus the projection factor, we construct $m$ KTs
of order two of the form $C_{ab}=L_{(a;b)}$ where $L_{a}=M_{a}+\eta_{a}.$

From condition (\ref{aa.2}) it follows that an AC $\eta^{a}$ satisfies
\[
\eta_{(a;bc)}=0
\]
therefore $C_{ab}=\eta_{\left(  a;b\right)  }$ defines the KT of order two. We
conclude that from the $s$ (say)\ proper ACs $n^{a}$ of a space we construct
the $s$ KTs $C_{ab}=\eta_{\left(  a;b\right)  }.$

\subsection{Killing tensors and collineations}

Concerning the KTs on a finite dimensional Riemannian space $V^{n}$ we have
the following results. Let $V^{n}$ be a Riemannian or Pseudo-Riemannian
manifold of dimension $n.$ Then:

\begin{enumerate}
\item The KTs of rank $m$ defined on $V^{n}$ form a linear space of dimension
less than or equal to%
\[
\frac{(n+m-1)!(n+m)!}{(n-1)!m!n!(m+1)!}%
\]
and the equality is attained if and only if $V^{n}$ is a space of constant
curvature \cite{kalnins}.

\item The maximum number of linearly independent Killing Tensors of order 2 in
a Riemannian manifold of dimension $n$ is $n(n+1)^{2}(n+2)/12$ and this is the
necessary and sufficient condition for the space to be of constant curvature
\cite{Thomas1946,Collinson1965,GlassGardinke}.

\item On flat spaces the Killing tensors are all reducible which means that
all are generated by products of Killing vectors \ \cite{Thomson1986A}.
\end{enumerate}

From the above results we conclude the following concerning the KTs of order 2
defined on $V^{n}.$

\begin{enumerate}
\item[a.] If the space admits $m$ gradient KVs $S_{J}$ $(J=1,2,...,m)$ and $r$
non-gradient KVs $M_{A}$ ($A=1,2,...,r)$ then we can construct $m^{2}%
+mr=m(m+r)~$KTs of rank two (excluding the metric), of the form $C_{ab}%
=L_{(a;b)}$ where
\begin{equation}
L_{a}=S_{I}S_{J,a}+S_{I}M_{Aa}. \label{FL.9a}%
\end{equation}

\item[b.] If the space admits another $k$ proper PCs $n^{a}$ with projection
factor $\phi$ and $k$ CKVs $G_{a}$ with conformal factor $-\phi$ then we may
construct another $k$ KTs of rank two of the same form $C_{ab}=L_{(a;b)}$
where%
\[
L_{a}=G_{a}+\eta_{a}.
\]
Finally if the space admits $r$ proper ACs then we may construct another $r$ KTs.
\end{enumerate}

We collect these results in the following

\begin{proposition}
\label{prop1}In a space $V^{n}$ the vector fields of the form
\begin{equation}
L_{a}=c_{1}KV_{a}+c_{2}PC_{a}+c_{3}AC_{a}+c_{1I}S_{I,a}+c_{2IJ}S_{I}%
S_{J,a}+2c_{IA}S_{I}M_{Aa} \label{FL.20}%
\end{equation}
where the proper PCs are CKVs such that the conformal factor is minus the
projective factor, the $AC_{a}$ are proper and $KV_{a}$ are the non-gradient
KVs produce KTs of order 2 of the form $C_{ab}=L_{(a;b)}$. In the case of
maximally symmetric spaces there do not exist proper PCs and proper ACs
therefore only the vectors generated by the KVs are necessary, that is, in
these spaces%
\begin{equation}
L_{a}=c_{1}KV_{a}+c_{1I}S_{I,a}+c_{2IJ}S_{I}S_{J,a}+2c_{3IA}S_{I}M_{Aa}.
\label{FL.21}%
\end{equation}
The $KVs$ give the solution $C_{ab}=0$ and the HV\ gives the solution
$C_{ab}=g_{ab}.$ These vectors construct all KTs $C_{ab}$ of a flat space. The
vectors $S_{I}S_{J,a}$ and $S_{I}M_{Aa}$ are gradient and non-gradient
non-proper ACs respectively.
\end{proposition}

It is important to note that in general not all KTs of order 2 in (a
non-flat)$\ \not V  ^{n}$ are reducible, that is of the form $C_{ab}%
=L_{(a;b)}.$

Let us apply the above proposition to the case of $E^{2}$ and reconstruct the
KTs. The $E^{2}~$~admits two gradient KVs $\partial_{x},\partial_{y}$ whose
generating functions are $x,y$ respectively and one non-gradient KV (the
rotation) $y\partial_{x}-x\partial_{y}$. From the KVs we can construct
$2(2+1)=6$ proper KTs. From (\ref{FL.21}) we have
\[
L_{a}=c_{11}\left(
\begin{array}
[c]{c}%
1\\
0
\end{array}
\right)  +c_{12}\left(
\begin{array}
[c]{c}%
0\\
1
\end{array}
\right)  +c_{211}x\left(
\begin{array}
[c]{c}%
1\\
0
\end{array}
\right)  +c_{212}x\left(
\begin{array}
[c]{c}%
0\\
1
\end{array}
\right)  +c_{221}y\left(
\begin{array}
[c]{c}%
1\\
0
\end{array}
\right)  +c_{222}y\left(
\begin{array}
[c]{c}%
0\\
1
\end{array}
\right)  +2c_{311}x\left(
\begin{array}
[c]{c}%
y\\
-x
\end{array}
\right)  +2c_{312}\left(
\begin{array}
[c]{c}%
y\\
-x
\end{array}
\right)
\]
or in a more convenient form
\begin{equation}
L_{a}=c_{11}\left(
\begin{array}
[c]{c}%
1\\
0
\end{array}
\right)  +c_{9}\left(
\begin{array}
[c]{c}%
0\\
1
\end{array}
\right)  +Ax\left(
\begin{array}
[c]{c}%
1\\
0
\end{array}
\right)  +By\left(
\begin{array}
[c]{c}%
0\\
1
\end{array}
\right)  +c_{8}y\left(
\begin{array}
[c]{c}%
1\\
0
\end{array}
\right)  +c_{10}x\left(
\begin{array}
[c]{c}%
0\\
1
\end{array}
\right)  +2ax\left(
\begin{array}
[c]{c}%
y\\
-x
\end{array}
\right)  -2\beta y\left(
\begin{array}
[c]{c}%
y\\
-x
\end{array}
\right)  \label{FL.14a}%
\end{equation}
which can be written%

\begin{equation}
L^{a}=\left(
\begin{array}
[c]{c}%
Ax+2axy+2\beta x^{2}+c_{8}y+c_{11}\\
By+2ay^{2}+2\beta xy+c_{10}x+c_{9}%
\end{array}
\right)  .\text{ } \label{FL.14}%
\end{equation}

The generic form of the tensor $C_{ab}=L_{(a;b)}$ in $E^{2}$ is%
\begin{equation}
C_{ab}=\left(
\begin{tabular}
[c]{ll}%
$L_{xx}$ & $\frac{1}{2}(L_{xy}+L_{yx})$\\
$\frac{1}{2}(L_{xy}+L_{yx})$ & $L_{yy}$%
\end{tabular}
\ \ \ \ \ \right)  =\left(
\begin{tabular}
[c]{ll}%
$A+2ay$ & $C-ax-\beta y$\\
$C-ax-\beta y$ & $B+2\beta y$%
\end{tabular}
\ \ \ \ \ \right)  \label{FL.14.1}%
\end{equation}
where \ for the constant $C$ we have the constraint $C=\frac{c_{8}+c_{10}}{2}$.

\section{Quadratic conservation laws and collineations}

\label{sec3}

Consider a dynamical system defined by the second-order differential equation%
\begin{equation}
\ddot{q}^{a}+\Gamma_{bc}^{a}\dot{q}^{b}\dot{q}^{c}+Q^{a}\left(  t,q^{c}%
\right)  =0. \label{fe.02}%
\end{equation}
where $\Gamma_{bc}^{a}\left(  q^{c}\right)  $ are the Christoffel symbols
defined by the kinetic metric (i.e. the kinetic energy)\ of the system. The
external forces $Q^{a}$ are given by
\begin{equation}
Q_{a}=V_{,a}-F_{a}\text{ } \label{fe.04}%
\end{equation}
where $V^{,a}$ stands for the total potential $V(t,q^{c})$ defined by the
conservative forces and $F^{a}$ stands for the sum of the non-conservative
forces. For this system we may consider the Lagrange function (assumed to be
regular )%
\begin{equation}
L\left(  t,q^{c},\dot{q}^{c}\right)  =\frac{1}{2}g_{ab}\left(  q^{c}\right)
\dot{q}^{a}\dot{q}^{b}-V(t,q^{c}). \label{fe.03}%
\end{equation}
and write (\ref{fe.02}) in the form $E_{a}\left(  L\right)  =Q_{a}$ where
$E_{a}=\frac{d}{dt}\frac{\partial}{\partial\dot{q}^{a}}-\frac{\partial
}{\partial q^{a}}$ is the Euler-Lagrange vector.

We consider next the function
\begin{equation}
I=K_{ab}(t,q^{c})\dot{q}^{a}\dot{q}^{b}+K_{a}(t,q^{c})\dot{q}^{a}+K(t,q^{c}),
\label{fe.01}%
\end{equation}
which is the quadratic in the velocities and demand that it is a first
integral, that is, satisfies the equation $\frac{dI}{dt}=0$ \ modulo the
equations of motion (\ref{fe.02}). This approach provides us the first
integrals of this form. This approach relies only on the dynamical equations
(\ref{fe.02}) and not on the Noether theorem which requires the (non-unique)
Lagrangian function. This idea is not new. It appears that is was raised for
the first time by Darboux \cite{darboux}\emph{\ }and later by Wittaker
\cite{Wittaker1}\emph{ }who both considered the case of Newtonian autonomous
holonomic systems with two degrees of freedom and determined most potentials
$V(q)$ for which these systems admit a quadratic first integral other than the
Hamiltonian (energy). The complete answer to this problem was given much later
by G. Thompson \cite{Thompson1}, \ for some extensions see also
\cite{dask1,ref001,ref002} and references therein.

The requirement $\frac{dI}{dt}=0$ leads to the equation
\begin{align}
0  &  =K_{(ab;c)}\dot{q}^{a}\dot{q}^{b}\dot{q}^{c}+\left(  K_{ab,t}%
+K_{a;b}\right)  \dot{q}^{a}\dot{q}^{b}\nonumber\\
&  +(K_{a,t}+K_{,a})\dot{q}^{a}-2K_{ar}\delta_{b}^{(r}F^{a)}\dot{q}%
^{b}\nonumber\\
&  +K_{,t}-K_{a}F^{a}.
\end{align}
from which using the various powers of the velocity follow the conditions%

\begin{equation}
K_{\left(  ab;c\right)  }=0 \label{FI.1.1a}%
\end{equation}

\begin{equation}
K_{(a;b)}=-K_{ab,t} \label{FI.1.3}%
\end{equation}%
\begin{equation}
K_{,b}+K_{b,t}-2K_{ab}F^{a}=0 \label{FI.1.4}%
\end{equation}%
\begin{equation}
K_{,t}=K_{a}F^{a}. \label{FI.1.5}%
\end{equation}

It is important to mention here, that these conditions reduce to the Noether
condition (\ref{FI.1}) when $F^{a}=0$ and $X=\left(  K_{b}^{a}\dot{q}%
^{b}+K^{a}\right)  \partial_{a}$ and to the weak Noether condition
\begin{equation}
X^{[1]}L+\frac{d\xi}{dt}L+\eta^{a}Q_{a}=\frac{df}{dt}. \label{fe.07}%
\end{equation}
which also known as Noether Bessel Hagen equation (NBH) \cite{Djukic1}
when~$F^{a}\neq0.$

The "solution\ "of conditions (\ref{FI.1.1a}) - (\ref{FI.1.5}) will provide
all quadratic first integrals of (\ref{fe.02}) for the form (\ref{fe.01}).

Condition (\ref{FI.1.1a}) implies that $K_{ab}$ is a KT for the kinetic metric
$\gamma_{ab}$. These conditions have been considered previously and they have
been "solved " for special cases. For example Kalotas et al.\cite{kalotas}
considered the case for $F^{a}=0$ and $K^{a}=0.$

On the other hand in \cite{p9} \ they have considered the case $K^{a}\neq0$,
and $K_{ab}=g\left(  t\right)  \gamma_{ab}$ for point symmetries, where
$K^{a}$ is a HV or a KV of the metric $\gamma_{ab}$ and $g_{,t}$ is the
homothetic factor of $K^{a}$.

In the general scenario considered here, $K_{ab}$ is a KT~of the form
$K_{ab}(t,q)=g(t)C_{ab}(q)$ where $C_{ab}(q^{a})$ is a general KT\ of the
kinetic metric. Then from (\ref{FI.1.3}) it follows that
\begin{equation}
K_{a}(t,q)=f(t)L_{a}(q)+B_{a}(q), \label{fe.08}%
\end{equation}
from where the symmetry conditions become%
\begin{equation}
f(t)L_{(a;b)}+B_{(a;b)}+g_{,t}C_{ab}=0, \label{fe.09}%
\end{equation}%
\begin{equation}
K_{,b}+f_{,t}L_{b}-2g(t)C_{ab}F^{a}=0, \label{fe.10}%
\end{equation}%
\begin{equation}
f(t)L_{a}F^{a}+B_{a}F^{a}-K_{,t}=0. \label{fe.11}%
\end{equation}

From the discussion of the previous section we can see that equation
(\ref{fe.09}) is again a geometric condition. It relates the KT $C_{ab}$ with
the vector fields~$L^{a}$ and $B^{b}$. The latter vector fields are given from
formulas (\ref{FL.20}) or (\ref{FL.21}) of Proposition \ref{prop1}.

We conclude that the quadratic conservation laws (\ref{fe.01}) with
nonvanishing linear term, exists if and only if the Riemannian space admits a
nontrivial projective algebra, which means that at least, a reducible KT
exists given by the geometric collineations as discussed in Section \ref{sec2}.

\subsubsection{Application}

\label{sec4}

In order to demonstrate the application of our geometric results let us
consider a dynamical system consists by the following differential equations%
\begin{equation}
\ddot{x}=F_{1}\left(  x,y\right)  ~~\text{and~~}\ddot{y}=F_{2}\left(
x,y\right)  \label{eq.veldep.10}%
\end{equation}
where $F_{1}\left(  x,y\right)  $,~$F_{2}\left(  x,y\right)  $ are arbitrary
functions. For the later system we observe that the underlying geometry is
that of the two-dimensional Euclidean space. Consider now, that conditions
(\ref{fe.09})-(\ref{fe.11}) are satisfied where $B^{a}=y\partial_{x}%
-x\partial_{y}$ and~$L^{a}=x\partial_{x}$. Hence from (\ref{fe.09}) it follows
that $C_{ab}=%
\begin{pmatrix}
c_{0} & 0\\
0 & 0
\end{pmatrix}
$ with $g_{,t}=\frac{1}{c_{0}}f$.\ 

Therefore, conditions (\ref{fe.10}), (\ref{fe.11}) provides%
\begin{equation}
K_{,x}+c_{0}g_{,tt}x-2g(t)F_{1}\left(  x,y\right)  =0,~~K_{,y}=0~~,
\label{fe.12}%
\end{equation}%
\begin{equation}
c_{0}g_{,t}\left(  F_{1}\left(  x,y\right)  \right)  +yF_{1}\left(
x,y\right)  -xF_{2}\left(  x,y\right)  -K_{,t}=0. \label{fe.13}%
\end{equation}

We assume that $g_{,t}=0$, that is $g\left(  t\right)  =\frac{1}{c_{0}}$, then
the latter system is simplified as
\begin{equation}
K_{,x}-\frac{2}{c_{0}}F_{1}\left(  x,y\right)  =0,~~K_{,y}=0~~, \label{fe.14}%
\end{equation}%
\begin{equation}
\left(  F_{1}\left(  x,y\right)  \right)  +yF_{1}\left(  x,y\right)
-xF_{2}\left(  x,y\right)  -K_{,t}=0. \label{fe.15}%
\end{equation}
which gives that $K=v\left(  x\right)  +n\left(  t\right)  $, and
\begin{equation}
F_{1}\left(  x,y\right)  =\frac{1}{2c_{0}}v_{,x}\left(  x\right)  \text{\ and
}F_{2}\left(  x,y\right)  =c_{0}\frac{yv\left(  x\right)  _{,x}}{2x}%
+\frac{n_{,t}\left(  t\right)  }{x}. \label{fe.16}%
\end{equation}

The latter means that all the dynamical systems of the form%
\begin{equation}
\ddot{x}=\frac{c_{0}}{2}v_{,x}\left(  x\right)  ~,~\ddot{y}=\frac
{c_{0}yv\left(  x\right)  _{,x}+2n_{,t}\left(  t\right)  }{2x} \label{fe.17}%
\end{equation}
admit the quadratic conservation law%
\begin{equation}
I=-\frac{1}{c_{0}}\dot{x}^{2}+y\dot{x}-x\dot{y}+v\left(  x\right)  +n\left(
t\right)  . \label{fe.18}%
\end{equation}

Working in the same way one is possible to classify all the types of dynamical
systems of the generic form (\ref{eq.veldep.10}) which admit quadratic
conservation laws and that without making use of Noether's theorem or other
type of symmetries but the collineations of the kinetic metric. Essentially
the solution of the equation $dI/dt=0$ is a matter of geometry.

\section{Conclusions}

\label{sec5}

The relation between the collineations of the kinetic metric and the existence
of first integrals for autonomous quasilinear differential equations has been
discussed. We reviewed previous results in the literature and derived a system
of equations whose solution derives all the quadratic first integrals they admit.

The purpose of our discussion was to make clear that first integrals for a
given dynamical system do not follow necessary from Noether's theorem but they
are rather a matter of geometry. Moreover, the characterization of some first
integrals as non-Noetherian is also meaningless.

In particular for a given dynamical system a geometry can be defined via the
kinetic metric and the existence of geometric collineations of the kinetic
metric is the only necessary and sufficient condition for first integrals to
exist. In a sense a dynamical system is somehow \textquotedblleft
constrained\textquotedblright\ by the geometry it creates, because the
collineations of this geometry provide the first integrals for the dynamical
and consequently specify its evolution.

The new mathematical relations which discussed in Section \ref{sec2} on the
construction of KTs with the use of the PCs indicate the importance of the
geometry for the existence of Lie/Noether symmetries. This means, that working
with collineations one is possible to extract information for the existence of
quadratic first integrals without performing the standard formal calculations
required by the Lie approach.

In a forthcoming work we plan to extend the geometric analysis on the symmetry
conditions (\ref{fe.09}) - (\ref{fe.11}) in order to understand in detail how
geometry is related to the force-term $F^{a}.$ Such an analysis will give us
important information for the geometric properties of integrable models and
also a possible geometric classification for the separable and non-separable
Hamiltonian systems, for instance see \cite{kabook}. Equally important step is
the general solution of the system of equations (\ref{fe.09}) - (\ref{fe.11})
which will provide all quadratic first integrals of (\ref{fe.02}) of the form
(\ref{fe.01}).

\bigskip

{\large {\textbf{Acknowledgements}}} \newline The research of AP was supported
by FONDECYT postdoctoral grant no. 3160121. AP thanks the University of Athens
for the hospitality provided this work was carried out.

\end{document}